\documentclass[twocolumn,showpacs,preprintnumbers,amsmath,amssymb]{revtex4}
\usepackage{graphicx}% Include figure files
\usepackage{dcolumn}% Align table columns on decimal point
\usepackage{bm}% bold math

\begin{document}

\preprint{APS/123-QED}
\title{Temporal aspects of one-dimensional completed scattering: An alternative view}

\author{N. L. Chuprikov}
\affiliation{Tomsk State Pedagogical University, 634041, Tomsk, Russia}
\altaffiliation[Also at ]{Physics Department, Tomsk State University.}

\date{\today}

\begin{abstract}
A {\it completed} scattering of a particle on a static one-dimensional (1D)
potential barrier is a combined quantum process to consist from two elementary
sub-processes (transmission and reflection) evolved coherently at all stages of
scattering and macroscopically distinct at the final stage. The existing model of
the process is clearly inadequate to its nature: all one-particle "observables" and
"tunneling times", introduced as quantities to be common for the sub-processes,
cannot be experimentally measured and, consequently, have no physical meaning; on
the contrary, quantities introduced for either sub-process have no basis, for the
time evolution of either sub-process is unknown in this model. We show that the wave
function to describe a completed scattering can be uniquely presented as the sum of
two solutions to the Schr\"odinger equation, which describe separately the
sub-processes at all stages of scattering. For symmetric potential barriers such
solutions are found explicitly. For either sub-process we define the time spent, on
the average, by a particle in the barrier region. We define it as the Larmor time.
As it turned out, this time is just Buttiker's dwell time averaged over the
corresponding localized state. Thus, firstly, we justify the known definition of the
local dwell time introduced by Hauge and co-workers as well by Leavens and Aers, for
now this time can be measured; secondly, we confirm that namely Buttiker's dwell
time gives the energy-distribution for the tunneling time; thirdly, we state that
all the definitions are valid only if they are based on the wave functions for
transmission and reflection found in our paper. Besides, we define the exact and
asymptotic group times to be auxiliary in timing the scattering process.
\end{abstract}

\maketitle

\newcommand{\Api}{A_{in}}
\newcommand{\Ami}{B_{in}}
\newcommand{\Apo}{A_{out}}
\newcommand{\Amo}{B_{out}}
\newcommand{\bpi}{a_{in}}
\newcommand{\bmi}{b_{in}}
\newcommand{\bpo}{a_{out}}
\newcommand{\bmo}{b_{out}}
\newcommand{\api}{a_{in}}
\newcommand{\ami}{b_{in}}
\newcommand{\apo}{a_{out}}
\newcommand{\amo}{b_{out}}

\section{Introduction} \label{aI}

For a long time scattering a particle on one-dimensional (1D) static potential
barriers have been considered in quantum mechanics as a representative of
well-understood phenomena. However, solving the so-called tunneling time problem
(TTP) (see reviews \cite{Ha2,La1,Olk,Ste,Mu0,Nu0,Ol3} and references therein) showed
that this is not the case.

At present there is a variety of approaches to introduce characteristic times for a
1D scattering. They are the group (Wigner) tunneling times (more known as the
"phase" tunneling times) \cite{Ha2,Wig,Har,Ha1,Ter}, different variants of the dwell
time \cite{Smi,Ja1,Ja2,Ha1,But,Le1,Nus,Go1,Mue,Bra}, the Larmor time
\cite{But,Baz,Ryb,Aer,Bu1,Lia,Zhi}, and the concept of the time of arrival which is
based on introducing either a suitable time operator (see, e.g.,
\cite{Aha,Mu4,Hah,Noh,Mu9}) or the positive operator valued measure (see review
\cite{Mu0}). A particular class of approaches to study the temporal aspects of a 1D
scattering includes the Bohmian \cite{Le2,Le3,Gru,Bo1,Kr1}, Feynman and Wigner ones
(see \cite{Sok,Yam,Ymm,Ya1,Kre} as well as \cite{La1,Mu0} and references therein).
One has also point out the papers \cite{Ga1,Ga2,Ga3} to study the characteristic
times of "the forerunner preceding the main tunneling signal of the wave created by
a source with a sharp onset".

As is known (see \cite{Ha2}), the main question of the TTP is that of the time
spent, on the average, by a particle in the barrier region in the case of a
completed scattering. Setting this problem implies that the particle's source and
detectors are located at a considerable distance from the potential barrier. The
answer to this question, for a given potential and initial state of a particle, is
evident must be unique. In particular, it must not depend on the details of
measurements with the removed detectors.

One has to recognize that the answer has not yet been found, and this elastic
scattering process looks at present like an unexplained phenomenon surrounded by
paradoxes. We bear in mind, in particular, 1) the lack of a causal relationship
between the transmitted and incident wave packets \cite{La2}; 2) a superluminal
propagation of a particle through opaque potential barriers (the Hartman effect)
\cite{Har,Mu6,Wi1,Ol1,So1}; 3) accelerating (on the average) a transmitted particle,
in the asymptotic region, as compared with an incident one \cite{La2}; 4) aligning
the average particle's spin with the magnetic field \cite{But,Lia}; 5) the Larmor
precession of the reflected particles under the non-zero magnetic field localized
beyond the barrier on the side of transmission \cite{Aer}.

At the first glance the Bohmian mechanics provides an adequate description of the
temporal aspects of a completed scattering (see, e.g., \cite{Le2,Le3,Gru,Bo1,Kr1}).
For its "causal" one-particle trajectories exclude, a priory, the appearance of the
above paradoxes. For example, the Hartman effect does not appear in this approach:
in the case of opaque rectangular barriers, the Bohmian dwell time, unlike Smith's
and Buttiker's dwell times, increases exponentially together with the barrier's
width (see also Section \ref{a4}).

It should be stressed however that the Bohmian model of a 1D completed scattering is
not free of paradoxes. As is well known, the region of location of the particle's
source consists in this model from two parts separated by some critical point. This
point is such that all particles starting from the sub-region, adjacent to the
barrier region, are transmitted by the barrier; otherwise they are reflected by it.
That is, the subensembles of transmitted and reflected particles are macroscopically
distinct in this model at all stages of scattering, what clearly contradicts the
main principles of quantum mechanics.

Note, the position of the critical point depends on the barrier's shape. For a
particle impinging the barrier from the left, this point approaches the left
boundary of the barrier when the latter becomes less transparent. Otherwise, the
critical point approaches minus infinity on the OX-axis. This property means, in
fact, that particles feel the barrier's shape, being however far from the barrier
region. Of course, this fact evidences, too, that the existing "causal" trajectories
of the Bohmian mechanics give an improper description of the scattering process.

From our viewpoint, all difficulties and paradoxes to arise in studying the temporal
aspects of a completed scattering result from the fact that setting this problem in
the existing framework of quantum mechanics is contradictory. On the one hand, in
the case of a completed scattering an observer deals only either with transmitted or
reflected particles, and, consequently, all one-particle observables must be
introduced individually for each sub-process. On the other hand, quantum mechanics,
as it stands, does not imply the introduction of observables for the sub-processes,
for its formalism does not provide the wave functions for transmission and
reflection, needed for computing the expectation values of observables.

So, in the case of a completed scattering, a conflicting situation arises already at
the stage of setting the problem: the nature of this process requires a separate
description of transmission and reflection; while quantum mechanics, as it stands,
does not allow such a description. This conflict underlies all controversy and
paradoxes to arise in solving the TTP: in fact, in the existing framework of quantum
theory, there are no observables which can be consistently introduced for this
process.

Note, this concerns not only characteristic times but also all observables to have
Hermitian operators. For example, averaging the particle's position and momentum
over the whole ensemble of particles does not give the expectation (i.e., most
probable) values of these quantities. As regards characteristic times, we have to
stress once more that among the existing time concepts neither separate nor common
times for transmission and reflection give the time spent by a particle in the
barrier region. In the first case, there is no basis to distinguish (theoretically
and experimentally) transmitted and reflected particles in the barrier region. In
the second case, characteristic times introduced cannot be properly interpreted
(see, e.g., discussion of the dwell and Larmor times in \cite{Ha1}); these times
describe neither transmitted nor reflected particles (ideal transmission and
reflection are exceptional cases).

At the same time there is a viewpoint that all the time scales introduced for a
completed scattering are valid: one has only to choose a suitable clock (operational
procedure) for each of them. This viewpoint is based on the assumption that timing a
quantum particle, without influencing the scattering process, is impossible in
principle. By this viewpoint the time measured should always depend on the clock
used for this purpose.

However, quantum phenomena, such as a completed scattering, have their own,
intrinsic spatial and temporal scales, and our main task is to learn to measure
these scales without influencing their values. In this paper we show that for the
problem under consideration this is possible. The above conflict can be resolved in
the framework of conventional quantum mechanics, and characteristic times for
transmission and reflection can be introduced. For measuring these time scales
without affecting the scattering process, one can exploit the Larmor precession of
the particle's spin under the infinitesimal magnetic field.

The plan of this paper is as follows. In (Section \ref{a0}) we introduce the concept
of combined and elementary quantum processes and states. By this concept, the state
of the whole quantum ensemble of particles, at the problem at hand, is a combined
one to represent a coherent superposition of two (elementary) states of the
(to-be-)transmitted and (to-be-)reflected subensembles of particles. In Section
\ref{a2} we present two solutions to the Schr\"odinger equation to describe
transmission and reflection at all stages of scattering. On their basis we define
the group, dwell and Larmor times for transmission and reflection (Section
\ref{a3}).

\section{The Schr\"odinger's cat paradox and 1D completed scattering: the concept of
combined and elementary states.} \label{a0}

For our purposes it is relevant to address the well-known Schrodinger's cat paradox
which displays explicitly a principal difference between macroscopically distinct
quantum states and their superpositions.

As is known, macroscopically distinct quantum states are symbolized in this paradox
by the 'dead-cat' and 'alive-cat' ones. Either may be associated with a single,
really existing cat which can be described in terms of one-cat observables. As
regards a superposition of these two states, it cannot be associated with a cat to
exist really (a cat cannot be dead and alive simultaneously). To calculate the
expectation values of one-cat observables for this state is evident to have no
physical sense.

As is known, quantum mechanics as it stands does not distinguish between the
'dead-cat' and 'alive-cat' states and their superposition. It postulates that all
its rules should be equally applied to macroscopically distinct states and their
superpositions. From our pint of view, the main lesson of the Schrodinger's cat
paradox is just that this postulate is erroneous. Quantum mechanics must distinguish
these two kinds of states on the conceptual level.

Hereinafter, any superposition of macroscopically distinct quantum states will be
referred to as a combined quantum state. All quantum states, like the "dead-cat" and
"alive-cat" ones, will be named here as elementary ones. Thereby we emphasize that
such states cannot be presented as a superposition of macroscopically distinct
states.

Note, the concepts of combined and elementary states are fully applicable to a 1D
completed scattering. Though we deal here with a microscopic object, at the final
stage of scattering the states of the subensembles of transmitted and reflected
particles are distinguished macroscopically. So that scattering a quantum particle
on the potential barrier is a combined process. It consists from two alternative
elementary one-particle sub-processes, transmission and reflection, evolved
coherently.

The main peculiarity of a time-dependent combined one-particle scattering state to
describe the combination of the two elementary sub-processes is that 1) in the
classical limit, such a state is associated with two one-particle trajectories,
rather than with one; 2) the squared modulus of such a state cannot be interpreted
as the probability density for one particle; 3) for this state it is meaningless to
calculate expectation values of one-particle observables, or to introduce
one-particle characteristic times and trajectories. All the quantum-mechanical rules
are applicable only to elementary states. Neglecting this circumstance leads to
paradoxes.

It is useful also to note that one has to distinguish between the interference of
different elementary states (e.g., the interference between the incident and
reflected waves in the case of a non-ideal reflection) and the self-interference of
the same elementary state (e.g., the interference between the incident and reflected
waves in the case of an ideal reflection). In the first case one deals with waves
which are not connected causally. In the second case, interfering waves are causally
connected with each other.

So, to explain properly a 1D completed scattering, we have to study the behavior of
the subensembles of transmitted and reflected particles at all stages of scattering.
At the first glance, this programm is impracticable in principle, since quantum
mechanics, as it stands, does not give the way of reconstructing the prehistory of
these subensembles by their final states. However, as will be shown below (see also
\cite{Ch5}), quantum mechanics implies such a reconstruction: we found two solutions
to the Schrodinger equation, which describe both the sub-processes at all stages of
scattering. Either consists from one incoming and only one outgoing (transmitted or
reflected) wave. Thus, though it is meaningless to say about to-be-transmitted or
to-be-reflected particles, the notions of to-be-transmitted and to-be-reflected
subensembles of particles are meaningful.

\newcommand{\ko}{\kappa_0^2}
\newcommand{\kj}{\kappa_j^2}
\newcommand{\kd}{\kappa_j d_j}
\newcommand{\kki}{\kappa_0\kappa_j}

\newcommand{\Ra}{R_{j+1}}
\newcommand{\Rb}{R_{(1,j)}}
\newcommand{\Rc}{R_{(1,j+1)}}

\newcommand{\Ta}{T_{j+1}}
\newcommand{\Tb}{T_{(1,j)}}
\newcommand{\Tc}{T_{(1,j+1)}}

\newcommand{\Wa}{w_{j+1}}
\newcommand{\Wb}{w_{(1,j)}}
\newcommand{\Wc}{w_{(1,j+1)}}

\newcommand{\UU}{u^{(+)}_{(1,j)}}
\newcommand{\VV}{u^{(-)}_{(1,j)}}

\newcommand{\ta}{t_{j+1}}
\newcommand{\tb}{t_{(1,j)}}
\newcommand{\tc}{t_{(1,j+1)}}

\newcommand{\tee}{\vartheta_{(1,j)}}

\newcommand{\tta}{\tau_{j+1}}
\newcommand{\ttb}{\tau_{(1,j)}}
\newcommand{\ttc}{\tau_{(1,j+1)}}

\newcommand{\FF}{\chi_{(1,j)}}
\newcommand {\aro}{(k)}
\newcommand {\da}{\partial}
\newcommand{\ppp}{\mbox{\hspace{5mm}}}
\newcommand{\ooo}{\mbox{\hspace{3mm}}}
\newcommand{\ooa}{\mbox{\hspace{1mm}}}

\section{Wave functions for transmission and reflection}\label{a2}
\subsection{Setting the problem for a 1D completed scattering} \label{a1}

Let us consider a particle incident from the left on the static potential barrier
$V(x)$ confined to the finite spatial interval $[a,b]$ $(a>0)$; $d=b-a$ is the
barrier width. Let its in-state, $\psi_{in}(x),$ at $t=0$ be a normalized function
to belong to the set $S_{\infty}$ consisting from infinitely differentiable
functions vanishing exponentially in the limit $|x|\to \infty$. The
Fourier-transform of such functions are known to belong to the set $S_{\infty},$
too. In this case the position, $\hat{x},$ and momentum, $\hat{p},$ operators both
are well-defined. Without loss of generality we will suppose that
\begin{eqnarray} \label{444}
<\psi_{in}|\hat{x}|\psi_{in}>=0,\ooo <\psi_{in}|\hat{p}|\psi_{in}> =\hbar k_0 >
0,\nonumber\\ <\psi_{in}|\hat{x}^2|\psi_{in}> =l_0^2;
\end{eqnarray}
here $l_0$ is the wave-packet's half-width at $t=0$ ($l_0<<a$).

We consider a completed scattering. This means that the average velocity, $\hbar
k_0/m,$ is large enough, so that the transmitted and reflected wave packets do not
overlap each other at late times. As for the rest, the relation of the average
energy of a particle to the barrier's height may be any by value.

We begin our analysis with the derivation of expressions for the incident,
transmitted and reflected wave packets to describe, in the problem at hand, the
whole ensemble of particles. For this purpose we will use the variant (see
\cite{Ch1}) of the well-known transfer matrix method \cite{Mez}. Let the wave
function $\psi_{full}(x,k)$ to describe the stationary state of a particle in the
out-of-barrier regions be written in the form
\begin{eqnarray} \label{1}
\psi_{full}(x;k)=e^{ikx}+b_{out}(k)e^{ik(2a-x)}, \ooo for \ooo x\le a;
\end{eqnarray}
\begin{eqnarray} \label{2}
\psi_{full}(x;k)=a_{out}(k)e^{ik(x-d)}, \ooo for \ooo x>b;
\end{eqnarray}
\noindent here $k=\sqrt{2mE}/\hbar;$ $E$ is the energy of a particle; $m$ is its
mass.

The coefficients entering this solution are connected by the transfer matrix ${\bf
Y}$:
\begin{eqnarray} \label{50}
\left(\begin{array}{c} 1 \\ b_{out}e^{2ika}
\end{array} \right)={\bf Y} \left(\begin{array}{c} a_{out}e^{-ikd} \\ 0
\end{array} \right), \ooa
{\bf Y}=\left(\begin{array}{cc} q & p \\ p^* & q^* \end{array} \right);
\end{eqnarray}
\newcommand{\iii}{\mbox{\hspace{10mm}}}
\begin{eqnarray} \label{500}
q=\frac{1}{\sqrt{T(k)}}\exp\left[i(kd-J(k))\right],\nonumber\\
p=\sqrt{\frac{R(k)}{T(k)}}\exp\left[i\left(\frac{\pi}{2}+ F(k)-ks\right)\right]
\end{eqnarray}
\noindent where $T$, $J$ and $F$ are the real tunneling parameters: $T(k)$ (the
transmission coefficient) and $J(k)$ (phase) are even and odd functions of $k$,
respectively; $F(-k)=\pi-F(k)$; $R(k)=1-T(k)$; $s=a+b$. We will suppose that the
tunneling parameters have already been calculated.

In the case of many-barrier structures, for this purpose one may use the recurrence
relations obtained in \cite{Ch1} just for these real parameters. For the rectangular
barrier of height $V_0$,
\begin{eqnarray} \label{501}
T=\left[1+\vartheta^2_{(+)}\sinh^2(\kappa d)\right]^{-1},\nonumber\\
J=\arctan\left(\vartheta_{(-)}\tanh(\kappa d)\right),\\F=0,\ooo
\kappa=\sqrt{2m(V_0-E)}/\hbar,\nonumber
\end{eqnarray}
if $E<V_0$; and
\begin{eqnarray} \label{502}
T=\left[1+\vartheta^2_{(-)}\sin^2(\kappa
d)\right]^{-1},\nonumber\\
J=\arctan\left(\vartheta_{(+)}\tan(\kappa d)\right),\\
F=\left\{\begin{array}{c} 0,\ooo if \ooo \vartheta_{(-)}\sin(\kappa d)\geq 0 \\
\pi,\ooo otherwise,
\end{array} \right.\nonumber\\
\kappa=\sqrt{2m(E-V_0)}/\hbar,\nonumber
\end{eqnarray}
if $E\geq V_0$; in both cases
$\vartheta_{(\pm)}=\frac{1}{2}\left(\frac{k}{\kappa}\pm \frac{\kappa}{k}\right)$
(see \cite{Ch1}).

Now, taking into account Exps. (\ref{50}) and (\ref{500}), we can write
in-asymptote, $\psi_{in}(x,t)$, and out-asymptote, $\psi_{out}(x,t)$, for the
time-dependent scattering problem (see \cite{Tei}):
\begin{eqnarray} \label{59}
\psi_{in}(x,t)=\frac{1}{\sqrt{2\pi}}\int_{-\infty}^{\infty} f_{in}(k,t)
e^{ikx}dk,\nonumber\\ f_{in}(k,t)=\Api(k) \exp[-i E(k)t/\hbar]
\end{eqnarray}
\begin{eqnarray} \label{60}
\psi_{out}(x,t)=\frac{1}{\sqrt{2\pi}}\int_{-\infty}^{\infty} f_{out}(k,t)
e^{ikx}dk,\nonumber\\ f_{out}(k,t)= f_{out}^{tr}(k,t)+f_{out}^{ref}(k,t)
\end{eqnarray}
\begin{eqnarray} \label{61}
f_{out}^{tr}(k,t)=\sqrt{T(k)}\Api(k) \exp[i(J(k)\nonumber\\-kd-E(k)t/\hbar)]
\end{eqnarray}
\begin{eqnarray} \label{62}
f_{out}^{ref}(k,t)=\sqrt{R(k)}\Api(-k)
\exp[-i(J(k)\nonumber\\-F(k)-\frac{\pi}{2}+2ka+E(k)t/\hbar)];
\end{eqnarray}
where Exps. (\ref{59}), (\ref{61}) and (\ref{62}) describe, respectively, the
incident, transmitted and reflected wave packets. Here $\Api(k)$ is the
Fourier-transform of $\psi_{in}(x).$ For example, for the Gaussian wave packet to
obey condition (\ref{444}), $\Api(k)=c\cdot \exp(-l_0^2(k-k_0)^2);$ $c$ is a
normalization constant.

\subsection{Incoming waves for transmission and reflection}\label{a21}

Let us now show that by the final states (\ref{60})-(\ref{62}) one can uniquely
reconstruct the prehistory of the subensembles of transmitted and reflected
particles at all stages of scattering. Let $\psi_{tr}$ and $\psi_{ref}$ be
searched-for wave functions for transmission (TWF) and reflection (RWF),
respectively. By our approach their sum should give the (full) wave function
$\psi_{full}(x,t)$ to describe the whole combined scattering process. From the
mathematical point of view our task is to find, for a particle impinging the barrier
from the left, such two solutions $\psi_{tr}$ and $\psi_{ref}$ to the Schr\"odinger
equation that, for any $t$,
\begin{equation} \label{261}
\psi_{full}(x,t)=\psi_{tr}(x,t)+\psi_{ref}(x,t);
\end{equation}
in the limit $t\to \infty,$
\begin{equation} \label{262}
\psi_{tr}(x,t)=\psi_{out}^{tr}(x,t), \ppp \psi_{ref}(x,t)=\psi_{out}^{ref}(x,t);
\end{equation}
where $\psi_{out}^{tr}(x,t)$ and $\psi_{out}^{ref}(x,t)$ are the transmitted and
reflected wave packets whose Fourier-transforms presented in (\ref{61}) and
(\ref{62}).

We begin with searching for the stationary wave functions for reflection,
$\psi_{ref}(x;k),$ and transmission, $\psi_{tr}(x;k).$ Let for $x\le a$
\begin{eqnarray} \label{265}
\psi_{ref}(x;k)=\Api^{ref}e^{ikx}+b_{out}e^{ik(2a-x)},\nonumber\\ \psi_{tr}(x;k)=
\Api^{tr}e^{ikx};
\end{eqnarray}
where $\Api^{tr}+\Api^{ref}=1.$

Since the RWF describes only reflected particles, which are expected to be absent
behind the barrier, the probability flux for $\psi_{ref}(x;k)$ should be equal to
zero -
\begin{eqnarray} \label{264}
|\Api^{ref}|^2-|b_{out}|^2=0.
\end{eqnarray}
In its turn, the probability flux for $\psi_{full}(x;k)$ and $\psi_{tr}(x;k)$ should
be the same -
\begin{eqnarray}\label{263}
|\Api^{tr}|^2=T(k)
\end{eqnarray}
Then, taking into account that $\psi_{tr}=\psi_{full}-\psi_{ref},$ we can exclude
$\psi_{tr}$ from Eq. (\ref{263}). As a result, we obtain
\begin{eqnarray} \label{2630}
\Re\left(\Api^{ref} \right)-|\amo|^2=0.
\end{eqnarray}
Since $|b_{out}|^2=R$, from Eqs. (\ref{264}) and (\ref{2630}) it follows that
$\Api^{ref}=\sqrt{R}(\sqrt{R}\pm i\sqrt{T}) \equiv \sqrt{R}\exp(i\lambda)$;
$\lambda=\pm\arctan(\sqrt{T/R})$.

So, a coherent superposition of the incoming waves to describe transmission and
reflection, for a given $E$, yields the incoming wave of unite amplitude, that
describes the whole ensemble of incident particles. In this case, not only
$\Api^{tr}+\Api^{ref}=1$, but also $|\Api^{tr}|^2+|\Api^{ref}|^2=1$! Besides, the
phase difference for the incoming waves to describe reflection and transmission
equals $\pi/2$ irrespective of the value of $E$.

Our next step is to show that only one root of $\lambda$ gives a searched-for
$\psi_{ref}(x;k).$ For this purpose the above solution should be extended into the
region $x>a$. To do this, we will restrict ourselves by symmetric potential
barriers, though the above derivation is valid for all barriers.

\subsection{Wave functions for transmission and reflection in the case of
symmetric potential barriers}\label{a22}

Let $V(x)$ be such that $V(x-x_c)=V(x_c-x);$ $x_c=(a+b)/2.$ As is known, for the
region of a symmetric potential barrier, one can always find odd, $u(x-x_c)$, and
even, $v(x-x_c)$, solutions to the Schr\"odinger equation. We will suppose here that
these functions are known. For example, for the rectangular potential barrier (see
Exps. (\ref{501}) and (\ref{502})),
\[ u(x)=\sinh(\kappa x),\ooo v(x)=\cosh(\kappa x),\ooo if\ooo E\le V_0;\]
\[ u(x)=\sin(\kappa x),\ooo v(x)=\cos(\kappa x),\ooo if\ooo E\ge V_0.\]
Note, $\frac{du}{dx}v-\frac{dv}{dx}u$ is a constant, which equals $\kappa$ in the
case of the rectangular barrier. Without loss of generality we will keep this
notation for any symmetric potential barrier.

Before finding $\psi_{ref}(x;k)$ and $\psi_{tr}(x;k)$ in the barrier region, we have
firstly to derive expressions for the tunneling parameters of symmetric barriers.
Let in the barrier region $\psi_{full}(x;k)=a_{full}\cdot u(x-x_c,k)+b_{full}\cdot
v(x-x_c,k).$ "Sewing" this expression together with Exps. (\ref{1}) and (\ref{2}) at
the points $x=a$ and $x=b$, respectively, we obtain
\begin{eqnarray*}
a_{full}=\frac{1}{\kappa}\left(P+P^*b_{out}\right)e^{ika}=
-\frac{1}{\kappa}P^*a_{out}e^{ika};\nonumber\\
b_{full}=\frac{1}{\kappa}\left(Q+Q^*b_{out}\right)e^{ika}=
\frac{1}{\kappa}Q^*a_{out}e^{ika};\nonumber
\end{eqnarray*}
\begin{eqnarray*}
Q=\left(\frac{du(x-x_c)}{dx}+i k u(x-x_c)\right)\Bigg|_{x=b};\nonumber\\
P=\left(\frac{dv(x-x_c)}{dx}+i k v(x-x_c)\right)\Bigg|_{x=b}.\nonumber
\end{eqnarray*}
As a result,
\begin{eqnarray} \label{300}
a_{out}=\frac{1}{2}\left(\frac{Q}{Q^*}-\frac{P}{P^*}\right);\ooa
b_{out}=-\frac{1}{2}\left(\frac{Q}{Q^*}+\frac{P}{P^*}\right).
\end{eqnarray}
As it follows from (\ref{50}), $a_{out}=\sqrt{T}\exp(iJ),$
$b_{out}=\sqrt{R}\exp\left(i\left(J-F-\frac{\pi}{2}\right)\right)$. Hence
$T=|a_{out}|^2,$ $R=|b_{out}|^2,$ $J=\arg(a_{out})$. Besides, for symmetric
potential barriers $F=0$ when $\Re(QP^*)>0$; otherwise, $F=\pi$.

Then, one can show that "sewing" the general solution $\psi_{ref}(x;k)$ in the
barrier region together with Exp. (\ref{265}) at $x=a$, for both the roots of
$\lambda$, gives odd and even functions in this region. For the problem considered,
only the former has a physical meaning. The corresponding roots for $\Api^{ref}$ and
$\Api^{tr}$ read as
\begin{eqnarray} \label{301}
\Api^{ref}=b_{out}\left(b^*_{out}-a^*_{out}\right);\ooa
\Api^{tr}=a^*_{out}\left(a_{out}+b_{out}\right)
\end{eqnarray}
One can easily show that in this case
\begin{eqnarray} \label{302}
\frac{Q^*}{Q}=-\frac{\Api^{ref}}{b_{out}}=\frac{\Api^{tr}}{a_{out}};
\end{eqnarray}
for $a\le x\le b$
\begin{eqnarray} \label{3000}
\psi_{ref}=\frac{1}{\kappa}\left(PA_{in}^{ref}+P^*b_{out}\right)e^{ika}u(x-x_c).
\end{eqnarray}
The extension of this solution onto the region $x\ge b$ gives
\begin{eqnarray*}
\psi_{ref}=-b_{out}e^{ik(x-d)}-\Api^{ref}e^{-ik(x-s)}.
\end{eqnarray*}

Let us now show that the searched for RWF is, in reality, zero to the right of the
barrier's midpoint. Indeed, as is seen from Exp. (\ref{3000}), $\psi_{ref}(x_c;k)=0$
for all values of $k$. In this case the probability flux, for any time-dependent
wave function formed from $\psi_{ref}(x;k)$, is equal to zero at the barrier's
midpoint for any value of time. This means that reflected particles impinging the
symmetric barrier from the left do not enter the region $x\ge x_c$. Thus,
$\psi_{ref}(x;k)\equiv 0$ for $x\ge x_c$. In the region $x\le x_c$ it is described
by Exps. (\ref{265}) and (\ref{3000}). For this solution, the probability density is
everywhere continuous and the probability flux is everywhere equal to zero.

As regards the searched-for TWF, one can easily show that
\begin{eqnarray} \label{303}
\psi_{tr}=a^l_{tr}u(x-x_c)+b_{tr}v(x-x_c)\ooa for\ooa a\le x\le x_c;
\end{eqnarray}
\begin{eqnarray} \label{304}
\psi_{tr}=a^r_{tr}u(x-x_c)+b_{tr}v(x-x_c)\ooa for\ooa x_c\le x\le b;
\end{eqnarray}
\begin{eqnarray} \label{305}
\psi_{tr}=a_{out}e^{ik(x-d)}\ooa for\ooa x\ge b.;
\end{eqnarray}
where
\begin{eqnarray*}
a^l_{tr}=\frac{1}{\kappa}PA_{in}^{tr}e^{ika},\ooa
b_{tr}=b_{full}=\frac{1}{\kappa}Q^*a_{out}e^{ika},\nonumber\\
a^r_{tr}=a_{full}=-\frac{1}{\kappa}P^*a_{out}e^{ika}\nonumber
\end{eqnarray*}
Like $\psi_{ref}(x;k),$ the TWF is everywhere continuous and the corresponding
probability flux is everywhere constant (we have to stress once more that this flux
has no discontinuity at the point $x=x_{c}$, though the first derivative of
$\psi_{tr}(x;k)$ on $x$ is discontinuous at this point). As in the case of the RWF,
wave packets formed from $\psi_{tr}(x;k)$ should evolve in time with a constant
norm.

So, for any value of $t$
\begin{eqnarray*}  {\bf T}=<\psi_{tr}(x,t)|\psi_{tr}(x,t)>=const;\nonumber\\ {\bf
R}=<\psi_{ref}(x,t)| \psi_{ref}(x,t)>=const;
\end{eqnarray*}
${\bf T}$ and ${\bf R}$ are the average transmission and reflection coefficients,
respectively. Besides,
\begin{eqnarray} \label{700100}
<\psi_{full}(x,t)|\psi_{full}(x,t)> ={\bf T}+{\bf R}=1.
\end{eqnarray}
From this it follows, in particular, that the scalar product of the wave functions
for transmission and reflection, $<\psi_{tr}(x,t)|\psi_{ref}(x,t)>,$ is a purely
imagine quantity to approach zero when $t\to\infty$.

\section{Characteristic times for transmission and reflection}\label{a3}

Now we are ready to proceed to the study of temporal aspects of a 1D completed
scattering. The wave functions for transmission and reflection presented in the
previous section permit us to introduce characteristic times for either sub-process.
Our main aim is to find, for each sub-process, the time spent, on the average, by a
particle in the barrier region. In doing so, we have to bear in mind that there may
be different approximations of this quantity. However, we have to remind that its
true value must not depend, for a completed scattering, on the choice of "clocks".

Measuring the tunneling time, under such conditions, implies that a particle has its
own, internal "clock" to remember the time spent by the particle in the spatial
region investigated. This means that the only way to measure the tunneling time for
a completed scattering is to exploit the internal degrees of freedom of quantum
particles. As is known, namely this idea underlies the Larmor-time concept based on
the Larmor precession of the particle's spin under the infinitesimal magnetic field.

In the above context, the concepts of the group and dwell times are rather auxiliary
ones, since they cannot be verified. Nevertheless, they may be useful for a better
understanding of the scattering process.

\subsection{Group times for transmission and reflection} \label{a31}

We begin our analysis from the group time concept to give the time spent by the
wave-packet's CM in the spatial regions. In other words, both for transmitted and
reflected particles, we begin with timing "mean-statistical particles" of these
subensembles (their motion is described by the Ehrenfest equations). In doing so, we
will distinguish exact and asymptotic group times.

\subsubsection{Exact group times} \label{a311}

\hspace*{\parindent} Let $t^{tr}_1$ and $t^{tr}_2$ be such moments of time that
\begin{equation} \label{80}
\frac{1}{{\bf T}}<\psi_{tr}(x,t^{tr}_1)|\hat{x}|\psi_{tr}(x,t^{tr}_1)> =a;
\end{equation}
\begin{equation} \label{81}
\frac{1}{{\bf T}}<\psi_{tr}(x,t^{tr}_2)|\hat{x}|\psi_{tr}(x,t^{tr}_2)> =b.
\end{equation}
\noindent Then, one can define the transmission time $\Delta t_{tr}(a,b)$ as the
difference $t^{tr}_2- t^{tr}_1$ where $t^{tr}_1$ is the smallest root of Eq.
(\ref{80}), and $t^{tr}_2$ is the largest root of Eq. (\ref{81}).

Similarly, for reflection, let $t_{(+)}$ and $t_{(-)}$ be such values of $t$ that
\begin{equation} \label{110}
\frac{1}{{\bf R}}<\psi_{ref}(x,t_{\pm})|\hat{x}|\psi_{ref}(x,t_{\pm})>=a,
\end{equation}
\noindent Then the exact group time for reflection, $\Delta t_{ref}(a,b),$ is
$\Delta t_{ref}(a,b)=t_{(+)}-t_{(-)}.$

Of course, a serious shortcoming of the exact characteristic times is that they fit
only for sufficiently narrow (in $x$-space) wave packets. For wide packets these
times give a very rough estimation of the time spent by a particle in the barrier
region. For example, one may a priory say that the exact group time for reflection,
for a sufficiently narrow potential barrier and/or wide wave packet, should be equal
to zero. In this case, the wave-packet's CM does not enter the barrier region.
\newcommand {\uta} {\tau_{tr}}
\newcommand {\utb} {\tau_{ref}}

\subsubsection{Asymptotic group times for transmission and reflection}\label{a312}

Note, the potential barrier influences a particle not only when its most probable
position is in the barrier region. For a completed scattering it is useful also to
introduce asymptotic group times to describe the passage of the particle in the
sufficiently large spatial interval $[a-L_1,b+L_2];$ where $L_1,L_2\gg l_0.$

It is evident that in this case, instead of the exact wave functions for
transmission and reflection, we may use the corresponding in- and out-asymptotes
derived in $k$-representation. The "full" in-asymptote, like the corresponding
out-asymptote, represents the sum of two wave packets:
\begin{eqnarray*}
f_{in}(k,t)=f_{in}^{tr}(k,t)+f_{in}^{ref}(k,t);
\end{eqnarray*}
\begin{eqnarray} \label{75}
f^{tr}_{in}(k,t)=\sqrt{T}\Api\exp[i(\lambda -\frac{\pi}{2} - E(k)t/\hbar)];
\end{eqnarray}
\begin{eqnarray} \label{76}
f^{ref}_{in}(k,t)=\sqrt{R}\Api\exp[i(\lambda- E(k)t/\hbar)];
\end{eqnarray}
$\lambda=\arg(A_{in}^{ref})$ (see (\ref{301})). One can easily show that
$|\lambda^\prime(k)|=\frac{|T^\prime|}{2\sqrt{R T}}$; hereinafter, the prime denotes
the derivative with respect to $k$.

For the average wave numbers in the asymptotic spatial regions we have
\[ <k>^{tr}_{in}=<k>^{tr}_{out}, \ppp <k>^{ref}_{in}=-<k>^{ref}_{out}.\]
Besides, at early and late times
\begin{eqnarray} \label{73}
<\hat{x}>^{tr}_{in}=\frac{\hbar t}{m}<k>^{tr}_{in} -<\lambda^\prime>^{tr}_{in};\\
<\hat{x}>^{tr}_{out}=\frac{\hbar t}{m}<k>^{tr}_{out}
-<J^\prime>^{tr}_{out}+d;\nonumber
\end{eqnarray}
\begin{eqnarray} \label{74}
<\hat{x}>^{ref}_{in}=\frac{\hbar t}{m}<k>^{ref}_{in}
-<\lambda^\prime>^{ref}_{in};\\
<\hat{x}>^{ref}_{out}=\frac{\hbar t}{m}<k>^{ref}_{out}
+<J^\prime-F^\prime>^{ref}_{out}+2a\nonumber
\end{eqnarray}
(henceforth, angle brackets denote averaging over the corresponding in- or
out-asymptotes).

As it follows from Exps. (\ref{73}) and (\ref{74}), the average starting points
$x_{start}^{tr}$ and $x_{start}^{ref}$, for the subensembles of transmitted and
reflected particles, respectively, read as
\begin{eqnarray} \label{730}
x_{start}^{tr}=-<\lambda^\prime>^{tr}_{in};\ooo x_{start}^{ref}=
-<\lambda^\prime>^{ref}_{in}.
\end{eqnarray}

The implicit assumption made in the standard wave-packet analysis is that
transmitted and reflected particles start, on the average, from the origin (in the
above setting the problem). However, by our approach, just $x_{start}^{tr}$ and
$x_{start}^{ref}$ are the average starting points of transmitted and reflected
particles, respectively. They are the initial values of $<\hat{x}>^{tr}_{in}$ and
$<\hat{x}>^{ref}_{in}$, which have the status of the expectation values of the
particle's position. They behave causally in time. As regards the average starting
point of the whole ensemble of particles, its coordinate is the initial value of
$<\hat{x}>_{in}$, which behaves non-causally in the course of scattering. This
quantity has no status of the {\it expectation} value of the particle's position.

Let us take into account Exps. (\ref{73}), (\ref{74}) and analyze the motion of a
particle in the spatial interval $[a-L_1,b+L_2]$. In particular, let us define the
transmission time for this region, making use the asymptotes of the TWF. We will
denote this time as $\Delta t^{as}_{tr}(a-L_1,b+L_2)$. The equations for the arrival
times $t^{tr}_1$ and $t^{tr}_2$ for the extreme points $x=a-L_1$ and $x=b+L_2$,
respectively, read as
\[ <\hat{x}>^{tr}_{in}(t^{tr}_1)=a-L_1;
\ppp<\hat{x}>^{tr}_{out}(t^{tr}_2)=b+L_2.
\]
\noindent Considering (\ref{73}), we obtain from here that the transmission time for
this interval is
\begin{eqnarray*}
\Delta t^{as}_{tr}(a-L_1,b+L_2)\equiv t^{tr}_2-t^{tr}_1\nonumber\\=\frac{m}{\hbar
<k>^{tr}_{in}}\left(<J^\prime>^{tr}_{out} -<\lambda^\prime>^{tr}_{in} +L_1+L_2
\right).\nonumber
\end{eqnarray*}
Similarly, for the reflection time $\Delta
t^{as}_{ref}(a-L_1,b+L_2),$ where $\Delta t_{ref}(a-L_1,b+L_2)=t^{ref}_2-t^{ref}_1$,
we have
\[ <\hat{x}>^{ref}_{in}(t^{ref}_1)=a-L_1,
\ppp<\hat{x}>^{ref}_{out}(t^{ref}_2)=a-L_1.
\]
\noindent Considering (\ref{74}), one can easily show that
\begin{eqnarray*}
\Delta t^{as}_{ref}(a-L_1,b+L_2)\equiv t^{ref}_2-t^{ref}_1\nonumber\\=\frac{m}{\hbar
<k>^{ref}_{in}}\left(<J^\prime - F^\prime>^{ref}_{out}-<\lambda^\prime>^{ref}_{in}
+2L_1\right).\nonumber
\end{eqnarray*}

The times $\uta^{as}$ ($\uta^{as}=\Delta t^{as}_{tr}(a,b)$) and $\utb^{as}$
($\utb^{as}=\Delta t^{as}_{ref}(a,b)$) are, respectively, the searched-for
asymptotic group times for transmission and reflection, for the barrier region:
\begin{eqnarray} \label{230}
\uta^{as}=\frac{m}{\hbar <k>^{tr}_{in}}\Big(<J^\prime>^{tr}_{out}
-<\lambda^\prime>^{tr}_{in}\Big),
\end{eqnarray}
\begin{eqnarray} \label{250}
\utb^{as}=\frac{m}{\hbar <k>^{ref}_{in}}\left(<J^\prime -
F^\prime>^{ref}_{out}-<\lambda^\prime>^{ref}_{in}\right)
\end{eqnarray}
Note, unlike the exact group times, the asymptotic ones may be negative by value:
they do not give the time spent by a particle in the barrier region (see also
Fig.1).

The lengths $d_{eff}^{tr}$ and $d_{eff}^{ref},$ where
\begin{eqnarray*}
d_{eff}^{tr}=<J^\prime>^{tr}_{out} -<\lambda^\prime>^{tr}_{in},\nonumber\\
d_{eff}^{ref}=<J^\prime-F^\prime>^{ref}_{out} -<\lambda^\prime>^{ref}_{in},\nonumber
\end{eqnarray*}
may be treated as the effective barrier's widths for transmission and
reflection, respectively.

\subsubsection{Average starting points and asymptotic group times for rectangular
potential barriers} \label{a313}

Let us consider the case of the rectangular barrier and obtain explicit expressions
for $d_{eff}(k)$ (now, both for transmission and reflection, $d_{eff}(k)=J^\prime(k)
-\lambda^\prime(k)$ since $F^\prime(k)\equiv 0$) which can be treated as the
effective width of the barrier for a particle with a given $k$. Besides, we will
obtain the corresponding expressions for the expectation value, $x_{start}(k)$, of
the staring point for this particle: $x_{start}(k)=-\lambda^\prime(k)$. It is
evident that in terms of $d_{eff}$ the above asymptotic times for a particle with
the well-defined momentum $\hbar k_0$ read as
\[ \uta^{as}=\utb^{as}=\frac{m d_{eff}(k_0)}{\hbar k_0}.\]

Using Exps. (\ref{501}) and (\ref{502}), one can show that, for the below-barrier
case ($E\le V_0$) -
\[ d_{eff}(k)=\frac{4}{\kappa}
\frac{\left[k^2+\kappa_0^2\sinh^2\left(\kappa d/2\right)\right]
\left[\kappa_0^2\sinh(\kappa d)-k^2 \kappa d\right]} {4k^2\kappa^2+
\kappa_0^4\sinh^2(\kappa d)}\]
\[ x_{start}(k)= -2\frac{\kappa_0^2}{\kappa}
\frac{(\kappa^2-k^2)\sinh(\kappa d)+k^2 \kappa d \cosh(\kappa d)} {4k^2\kappa^2+
\kappa_0^4\sinh^2(\kappa d)};\] for the above-barrier case ($E\ge V_0)$
-
\[ d_{eff}(k)=\frac{4}{\kappa} \frac{\left[k^2-\beta
\kappa_0^2\sin^2\left(\kappa d/2\right)\right]\left[k^2 \kappa d-\beta
\kappa_0^2\sin(\kappa d)\right]} {4k^2\kappa^2+\kappa_0^4\sin^2(\kappa d)}\]
\[ x_{start}(k)= -2\beta \frac{\kappa_0^2}{\kappa} \cdot
\frac{(\kappa^2+k^2)\sin(\kappa d)-k^2 \kappa d \cos(\kappa d)} {4k^2\kappa^2+
\kappa_0^4\sin^2(\kappa d)},\] where $\kappa_0=\sqrt{2m|V_0|/\hbar^2};$ $\beta=1$,
if $V_0>0$; otherwise, $\beta=-1.$

Note, $d_{eff}\to d$ and $x_{start}(k) \to 0$, in the limit $k\to \infty$. For
infinitely narrow in $x$-space wave packets, this property ensures the coincidence
of the average starting points for both subensembles with that for all particles.
For wide barriers, when  $\kappa d\gg 1$ and $E\le V_0$, we have $d_{eff}\approx
2/\kappa$ and $x_{start}(k)\approx 0.$ That is, the asymptotic group transmission
time saturates with increasing the width of an opaque potential barrier.

It is important to stress that for the $\delta$-potential, $V(x)= W \delta(x-a),$
$d_{eff}\equiv 0$. The subensembles of transmitted and reflected particles start, on
the average, from the point $x_{start}(k):$ $x_{start}(k)=-2m\hbar^2 W/(\hbar^4
k^2+m^2W^2).$

\subsection{Dwell times}\label{a32}

Let us now consider the stationary scattering problem. It describes the limiting
case of a scattering of wide wave packets, when the group-time concept leads to a
large error in timing a particle.

\subsubsection{Dwell time for transmission} \label{a321}

Note, in the case of transmission the density of the probability flux, $I_{tr}$, for
$\psi_{tr}(x;k)$ is everywhere constant and equal to $T\cdot\hbar k/m$. The
velocity, $v_{tr}(x,k)$, of an infinitesimal element of the flux, at the point $x,$
equals $v_{tr}(x)=I_{tr}/|\psi_{tr}(x;k)|^2.$ Outside the barrier region the
velocity is everywhere constant: $v_{tr}=\hbar k/m$. In the barrier region it
depends on $x$. In the case of an opaque rectangular potential barrier, $v_{tr}(x)$
decreases exponentially when the infinitesimal element approaches the midpoint
$x_c$. One can easily show that $|\psi_{tr}(a;k)|=|\psi_{tr}(b;k)|=\sqrt{T}$, but
$|\psi_{tr}(x_c;k)|\sim\sqrt{T}\exp(\kappa d/2)$.

Thus, any selected infinitesimal element of the flux passes the barrier region for
the time $\tau^{tr}_{dwell}$, where
\begin{eqnarray} \label{4005}
\tau^{tr}_{dwell}(k)=\frac{1}{I_{tr}}\int_a^b|\psi_{tr}(x;k)|^2 dx.
\end{eqnarray}
By analogy with \cite{But} we will call this time scale the dwell time for
transmission.

For the rectangular barrier this time reads (for $E< V_0$ and $E\ge V_0$,
respectively) as
\begin{eqnarray} \label{4007}
\tau^{tr}_{dwell}=\frac{m}{2\hbar k\kappa^3}\left[\left(\kappa^2-k^2\right)\kappa d
+\kappa_0^2 \sinh(\kappa d)\right],
\end{eqnarray}
\begin{eqnarray} \label{4009}
\tau^{tr}_{dwell}=\frac{m}{2\hbar k\kappa^3}\left[\left(\kappa^2+k^2\right)\kappa d
-\beta \kappa_0^2 \sin(\kappa d)\right].
\end{eqnarray}

\subsubsection{Dwell time for reflection} \label{a322}

In the case of reflection the situation is less simple. The above arguments are not
applicable here, for the probability flux for $\psi_{ref}(x,k)$ is zero. However, as
is seen, the dwell time for transmission coincides, in fact, with Buttiker's dwell
time introduced however on the basis of the wave function for transmission.
Therefore, making use of the arguments by Buttiker, let us define the dwell time for
reflection, $\tau^{ref}_{dwell}$, as
\begin{eqnarray} \label{40014}
\tau^{ref}_{dwell}(k)=\frac{1}{I_{ref}} \int_a^{x_c}|\psi_{ref}(x,k)|^2 dx;
\end{eqnarray}
where $I_{ref}=R\cdot \hbar k/m$ is the incident probability flux for reflection.

Again, for the rectangular barrier
\begin{eqnarray} \label{40030}
\tau^{ref}_{dwell}=\frac{m k}{\hbar \kappa}\cdot\frac{\sinh(\kappa d)-\kappa
d}{\kappa^2+\kappa^2_0 \sinh^2(\kappa d/2)}\ooo for \ooo E<V_0;
\end{eqnarray}
\begin{eqnarray} \label{40031}
\tau^{ref}_{dwell}=\frac{m k}{\hbar \kappa}\cdot\frac{\kappa d-\sin(\kappa
d)}{\kappa^2+\beta\kappa^2_0 \sin^2(\kappa d/2)}\ooo for \ooo E\ge V_0.
\end{eqnarray}
As is seen, for rectangular barriers the dwell times for transmission and reflection
do not coincide with each other, unlike the asymptotic group times.

We have to stress once more that Exps. (\ref{4005}) and (\ref{40014}), unlike
Smith's, Buttiker's and Bohmian dwell times, are defined in terms of the TWF and
RWF. As will be seen from the following, the dwell times introduced can be justified
in the framework of the Larmor-time concept.

\subsection{Larmor times for transmission and reflection}\label{a33}

As was said above, both the group and dwell time concepts do not give the way of
measuring the time spent by a particle in the barrier region. This task can be
solved in the framework of the Larmor time concept. As is known, the idea to use the
Larmor precession as clocks was proposed by Baz' \cite{Baz} and developed later by
Rybachenko \cite{Ryb} and B\"{u}ttiker \cite{But} (see also \cite{Aer,Lia}). However
the known concept of Larmor time has a serious shortcoming. It was introduced in
terms of asymptotic values (see \cite{But,Aer,Lia}). In this connection, our next
step is to define the Larmor times for transmission and reflection, taking into
account the expressions for the corresponding wave functions in the barrier region.

\subsubsection{Preliminaries} \label{a330}

Let us consider the quantum ensemble of electrons moving along the $x$-axis and
interacting with the symmetrical time-independent potential barrier $V(x)$ and small
magnetic field (parallel to the $z$-axis) confined to the finite spatial interval
$[a,b].$ Let this ensemble be a mixture of two parts. One of them consists from
electrons with spin parallel to the magnetic field. Another is formed from particles
with antiparallel spin.

Let at $t=0$ the in state of this mixture be described by the spinor
\begin{eqnarray} \label{9001}
\Psi_{in}(x)=\frac{1}{\sqrt{2}}\left(\begin{array}{c} 1 \\ 1
\end{array} \right)\psi_{in}(x),
\end{eqnarray}
where $\psi_{in}(x)$ is a normalized function to satisfy conditions (\ref{444}). So
that we will consider the case, when the spin coherent in state (\ref{9001}) is the
eigenvector of $\sigma_x$ with the eigenvalue 1 (the average spin of the ensemble of
incident particles is oriented along the $x$-direction); hereinafter, $\sigma_x,$
$\sigma_y$ and $\sigma_z$ are the Pauli spin matrices.

For electrons with spin up (down), the potential barrier effectively decreases
(increases), in height, by the value $\hbar\omega_L/2$; here $\omega_L$ is the
frequency of the Larmor precession; $\omega_L=2\mu B/\hbar,$ $\mu$ denotes the
magnetic moment. The corresponding Hamiltonian has the following form,
\begin{eqnarray} \label{900200}
\hat{H}=\frac{\hat{p}^2}{2m}+V(x)-\frac{\hbar\omega_L}{2}\sigma_z, \ooo if\ooo
x\in[a,b];\nonumber\\ \hat{H}=\frac{\hat{p}^2}{2m}, \ooo otherwise.
\end{eqnarray}
For $t>0$, due to the influence of the magnetic field, the states of particles with
spin up and down become different. The probability to pass the barrier is different
for them. Let for any value of $t$ the spinor to describe the state of particles
read as
\begin{eqnarray} \label{9002}
\Psi_{full}(x,t)=\frac{1}{\sqrt{2}}\left(\begin{array}{c}
\psi_{full}^{(\uparrow)}(x,t) \\
\psi_{full}^{(\downarrow)}(x,t) \end{array} \right).
\end{eqnarray}

In accordance with (\ref{261}), either spinor component can be uniquely presented as
a coherent superposition of two probability fields to describe transmission and
reflection:
\begin{eqnarray} \label{9003}
\psi_{full}^{(\uparrow\downarrow))}(x,t)=
\psi_{tr}^{(\uparrow\downarrow))}(x,t)+\psi_{ref}^{(\uparrow\downarrow))}(x,t);
\end{eqnarray}
note that $\psi_{ref}^{(\uparrow\downarrow)}(x,t)\equiv 0$ for $x\ge x_c$. As a
consequence, the same decomposition takes place for spinor (\ref{9002}):
$\Psi_{full}(x,t)= \Psi_{tr}(x,t)+\Psi_{ref}(x,t).$

We will suppose that all the wave functions for transmission and reflection are
known. It is important to stress here (see (\ref{700100}) that
\begin{eqnarray} \label{900100}
<\psi_{full}^{(\uparrow\downarrow)}(x,t)|\psi_{full}^{(\uparrow\downarrow)}(x,t)>
=T^{(\uparrow\downarrow)}+R^{(\uparrow\downarrow)}=1;\\
T^{(\uparrow\downarrow)}=<\psi_{tr}^{(\uparrow\downarrow)}(x,t)|
\psi_{tr}^{(\uparrow\downarrow)}(x,t)>=const;\nonumber\\
R^{(\uparrow\downarrow)}=<\psi_{ref}^{(\uparrow\downarrow)}(x,t)|
\psi_{ref}^{(\uparrow\downarrow)}(x,t)>=const;\nonumber
\end{eqnarray}
$T^{(\uparrow\downarrow)}$ and $R^{(\uparrow\downarrow)}$ are the (real)
transmission and reflection coefficients, respectively, for particles with spin up
$(\uparrow)$ and down $(\downarrow)$. Let further
$T=(T^{(\uparrow)}+T^{(\downarrow)})/2$ and $R=(R^{(\uparrow)}+R^{(\downarrow)})/2$
be quantities to describe all particles.

\subsubsection{Time evolution of the spin polarization of particles} \label{a332}

To study the time evolution of the average particle's spin, we have to find the
expectation values of the spin projections $\hat{S}_x$, $\hat{S}_y$ and $\hat{S}_z$.
Note, for any $t$
\begin{eqnarray*}
<\hat{S}_x>_{full}\equiv \frac{\hbar}{2}\sin(\theta_{full})\cos(\phi_{full})\\=\hbar
\cdot \Re(<\psi_{full}^{(\uparrow)}|\psi_{full}^{(\downarrow)}>);
\end{eqnarray*}
\begin{eqnarray} \label{9006}
 <\hat{S}_y>_{full}\equiv
\frac{\hbar}{2}\sin(\theta_{full})\sin(\phi_{full})=\nonumber\\ \hbar\cdot
\Im(<\psi_{full}^{(\uparrow)}|\psi_{full}^{(\downarrow)}>);
\end{eqnarray}
\begin{eqnarray*}
<\hat{S}_z>_{full}\equiv \frac{\hbar}{2}\cos(\theta_{full})\\=\frac{\hbar}{2}
\left[<\psi_{full}^{(\uparrow)}|\psi_{full}^{(\uparrow)}>
-<\psi_{full}^{(\downarrow)}|\psi_{full}^{(\downarrow)}>\right].
\end{eqnarray*}
Similar expressions are valid for transmission and reflection:
\begin{eqnarray*}
<\hat{S}_x>_{tr}=\frac{\hbar}{T}
\Re(<\psi_{tr}^{(\uparrow)}|\psi_{tr}^{(\downarrow)}>),\\
<\hat{S}_y>_{tr}=\frac{\hbar}{T}
\Im(<\psi_{tr}^{(\uparrow)}|\psi_{tr}^{(\downarrow)}>),\\
<\hat{S}_z>_{tr}=\frac{\hbar}{2T}
\Big(<\psi_{tr}^{(\uparrow)}|\psi_{tr}^{(\uparrow)}>
-<\psi_{tr}^{(\downarrow)}|\psi_{tr}^{(\downarrow)}>\Big),
\end{eqnarray*}
\begin{eqnarray*}
<\hat{S}_x>_{ref}=\frac{\hbar}{R}
\Re(<\psi_{ref}^{(\uparrow)}|\psi_{ref}^{(\downarrow)}>),\\
<\hat{S}_y>_{ref}=\frac{\hbar}{R}
\Im(<\psi_{ref}^{(\uparrow)}|\psi_{ref}^{(\downarrow)}>),\\
<\hat{S}_z>_{ref}=\frac{\hbar}{2R}
\left(<\psi_{ref}^{(\uparrow)}|\psi_{ref}^{(\uparrow)}>
-<\psi_{ref}^{(\downarrow)}|\psi_{ref}^{(\downarrow)}>\right).
\end{eqnarray*}

Note, $\theta_{full}=\pi/2,$ $\phi_{full}=0$ at $t=0.$ However, this is not the case
for transmission and reflection. Namely, for $t=0$ we have
\begin{eqnarray*}
\phi_{tr,ref}^{(0)}=\arctan\left(\frac{\Im(<\psi_{tr,ref}^{(\uparrow)}(x,0)|
\psi_{tr,ref}^{(\downarrow)}(x,0)>)}
{\Re(<\psi_{tr,ref}^{(\uparrow)}(x,0)|\psi_{tr,ref}^{(\downarrow)}(x,0)>)}\right);
\end{eqnarray*}
\begin{eqnarray*}
\theta_{tr,ref}^{(0)}=\arccos\Big(<\psi_{tr,ref}^{(\uparrow)}(x,0)|
\psi_{tr,ref}^{(\uparrow)}(x,0)>\\
-<\psi_{tr,ref}^{(\downarrow)}(x,0)|\psi_{tr,ref}^{(\downarrow)}(x,0)>\Big);
\end{eqnarray*}

Since the norms of $\psi_{tr}^{(\uparrow\downarrow)}(x,t)$ and
$\psi_{ref}^{(\uparrow\downarrow)}(x,t)$ are constant,
$\theta_{tr}(t)=\theta_{tr}^{(0)}$ and $\theta_{ref}(t)=\theta_{ref}^{(0)}$ for any
value of $t$. For the $z$-components of spin we have
\begin{eqnarray} \label{90018}
<\hat{S}_z>_{tr}(t)=\hbar\frac{T^{(\uparrow)}-
T^{(\downarrow)}}{T^{(\uparrow)}+T^{(\downarrow)}},\nonumber\\
<\hat{S}_z>_{ref}(t)=\hbar\frac{R^{(\uparrow)}-
R^{(\downarrow)}}{R^{(\uparrow)}+R^{(\downarrow)}}.
\end{eqnarray}

So, since the operator $\hat{S}_z$ commutes with Hamiltonian (\ref{900200}), this
projection of the particle's spin should be constant, on the average, both for
transmission and reflection. From the most beginning the subensembles of transmitted
and reflected particles possess a nonzero average $z$-component of spin (though it
equals zero for the whole ensemble of particles, for the case considered) to be
conserved in the course of scattering. By our approach it is meaningless to use the
angles $\theta_{tr}^{(0)}$ and $\theta_{ref}^{(0)}$ as a measure of the time spent
by a particle in the barrier region.

\subsubsection{Larmor precession caused by the infinitesimal magnetic field confined to
the barrier region} \label{a333}

As in \cite{But,Lia}, we will suppose further that the applied magnetic field is
infinitesimal. In order to introduce characteristic times let us find the
derivations $d\phi_{tr}/dt$ and $d\phi_{ref}/dt.$ For this purpose we will use the
Ehrenfest equations for the average spin of particles:
\begin{eqnarray*}
\frac{d<\hat{S}_x>_{tr}}{dt}=-\frac{\hbar\omega_L}{T} \int_a^b
\Im[(\psi_{tr}^{(\uparrow)}(x,t))^*\psi_{tr}^{(\downarrow)}(x,t)]dx\\
\frac{d<\hat{S}_y>_{tr}}{dt}=\frac{\hbar\omega_L}{T} \int_a^b
\Re[(\psi_{tr}^{(\uparrow)}(x,t))^*\psi_{tr}^{(\downarrow)}(x,t)]dx\\
\frac{d<\hat{S}_x>_{ref}}{dt}=-\frac{\hbar\omega_L}{R} \int_a^{x_c}
\Im[(\psi_{ref}^{(\uparrow)}(x,t))^*\psi_{ref}^{(\downarrow)}(x,t)]dx\\
\frac{d<\hat{S}_y>_{ref}}{dt}=\frac{\hbar\omega_L}{R} \int_a^{x_c}
\Re[(\psi_{ref}^{(\uparrow)}(x,t))^*\psi_{ref}^{(\downarrow)}(x,t)]dx.
\end{eqnarray*}
Note, $\phi_{tr}= \arctan\left(<\hat{S}_y>_{tr}/<\hat{S}_x>_{tr}\right),$
$\phi_{ref}=\arctan\left(<\hat{S}_y>_{ref}/<\hat{S}_x>_{ref}\right)$ Hence, in the
case of infinitesimal magnetic field and chosen initial conditions, when
$|<\hat{S}_y>_{tr,ref}|\ll|<\hat{S}_x>_{tr,ref}|,$ we have
\begin{eqnarray*}
\frac{d \phi_{tr}}{dt}=\frac{1}{<\hat{S}_x>_{tr}}\cdot
\frac{d<\hat{S}_y>_{tr}}{dt};\\ \frac{d
\phi_{ref}}{dt}=\frac{1}{<\hat{S}_x>_{ref}}\cdot \frac{d<\hat{S}_y>_{ref}}{dt}.
\end{eqnarray*}
Then, considering the above expressions for the spin projections and their
derivatives on $t$, we obtain
\[\frac{d \phi_{tr}}{dt}=\omega_L \frac{\int_a^b
\Re[(\psi_{tr}^{(\uparrow)}(x,t))^*\psi_{tr}^{(\downarrow)}(x,t)]dx}
{\int_{-\infty}^\infty
\Re[(\psi_{tr}^{(\uparrow)}(x,t))^*\psi_{tr}^{(\downarrow)}(x,t)]dx};\] \[\frac{d
\phi_{ref}}{dt}=\omega_L \frac{\int_a^{x_c}
\Re[(\psi_{ref}^{(\uparrow)}(x,t))^*\psi_{ref}^{(\downarrow)}(x,t)]dx}
{\int_{-\infty}^{x_c}
\Re[(\psi_{ref}^{(\uparrow)}(x,t))^*\psi_{ref}^{(\downarrow)}(x,t)]dx}.\] Or, taking
into account that in the first order approximation on $\omega_L$, when
$\psi_{tr}^{(\uparrow)}(x,t)=\psi_{tr}^{(\downarrow)}(x,t)=\psi_{tr}(x,t)$ and
$\psi_{ref}^{(\uparrow)}(x,t)= \psi_{ref}^{(\downarrow)}(x,t)=\psi_{ref}(x,t),$ we
have
\[\frac{d \phi_{tr}}{dt}\approx\frac{\omega_L}{{\bf T}} \int_a^b
|\psi_{tr}(x,t)|^2dx;\] \[\frac{d \phi_{ref}}{dt}\approx\frac{\omega_L}{{\bf R}}
\int_a^{x_c} |\psi_{ref}(x,t)|^2dx;\] note, in this limit, $T\to{\bf T}$ and
$R\to{\bf R}$.

As is supposed in our setting the problem, both at the initial and final instants of
time, a particle does not interact with the potential barrier and magnetic field. In
this case, without loss of exactness, the angles of rotation ($\Delta\phi_{tr}$ and
$\Delta\phi_{ref}$) of spin under the magnetic field, in the course of a completed
scattering, can be written in the form,
\begin{eqnarray} \label{90020}
\Delta\phi_{tr}=\frac{\omega_L}{{\bf T}} \int_{-\infty}^\infty dt \int_a^b
dx|\psi_{tr}(x,t)|^2,\nonumber\\ \Delta\phi_{ref}=\frac{\omega_L}{{\bf R}}
\int_{-\infty}^\infty dt \int_a^{x_c} dx|\psi_{ref}(x,t)|^2.
\end{eqnarray}
On the other hand, both the quantities can be written in the form:
$\Delta\phi_{tr}=\omega_L \tau^L_{tr}$ and $\Delta\phi_{eef}=\omega_L \tau^L_{ref},$
where $\tau^L_{tr}$ and $\tau^L_{ref}$ are the Larmor times for transmission and
reflection. Comparing these expressions with (\ref{90020}), we eventually obtain
\begin{eqnarray} \label{922}
\tau^L_{tr}=\frac{1}{{\bf T}} \int_{-\infty}^\infty dt \int_a^b
dx|\psi_{tr}(x,t)|^2, \nonumber\\
\tau^L_{ref}=\frac{1}{{\bf R}} \int_{-\infty}^\infty dt \int_a^{x_c}
dx|\psi_{ref}(x,t)|^2.
\end{eqnarray}
These are just the searched-for definitions of the Larmor times for transmission and
reflection.

As is seen, if the state of a particle is described by a normalized wave function
$\psi$, then the time $\tau(\psi;a,b)$ spent by the particle in the barrier region
is
\begin{eqnarray} \label{822}
\tau(\psi;a,b)=\int_{-\infty}^\infty dt \int_a^b dx|\psi(x,t)|^2.
\end{eqnarray}
This definition is just that introduced in \cite{Ha1,Ja1,Le1,Ha2}) on the basis of
classical mechanics (see also \cite{Nu0,Ol3,Mue}); note that in both cases the
integrals are calculated over the whole completed scattering. Thus, on the one hand,
our approach justifies the definition (B2), since this expression is obtained now as
the Larmor time. As a consequence, it can be verified experimentally. On the other
hand, we correct the domain of the validity of this expression. By our approach, it
is meaningful only in the framework of the separate description of transmission and
reflection, based on the solutions $\psi_{tr}(x,t)$ and $\psi_{ref}(x,t)$ found
first in the present paper.

Our next step is to transform Exps. (\ref{922}). Note, for transmission,
$\psi_{tr}(x,t)$ reads as
\begin{eqnarray*}
\psi_{tr}(x,t)=\frac{1}{\sqrt{2\pi}}\int_{-\infty}^{\infty}
A_{in}(k)\psi_{tr}(x,k)e^{-i E(k)t/\hbar}dk;
\end{eqnarray*}
where $\psi_{tr}(x,k)$ is the stationary wave function for transmission (see Section
\ref{a2}). Then the integral $I=\int_{-\infty}^\infty dt \int_a^b
dx|\psi_{tr}(x,t)|^2$ in (\ref{922}) can be reduced, by integrating on $t$, to the
form
\begin{eqnarray*}
I=\frac{\hbar}{\pi}\int_{-\infty}^{\infty} dk^\prime dk A_{in}^*(k^\prime)
A_{in}(k)\int_a^b dx \psi^*_{tr}(x,k^\prime) \psi_{tr}(x,k)\\
\times\lim_{\Delta t\to\infty} \frac{\sin[(E(k^\prime)-E(k))\Delta
t/\hbar]}{E(k^\prime)-E(k)}
\end{eqnarray*}
However,
\begin{eqnarray*}
\lim_{\Delta t\to\infty} \frac{\sin[(E(k^\prime)-E(k))\Delta
t/\hbar]}{E(k^\prime)-E(k)}=\frac{\pi}{\hbar}\delta[(E(k^\prime)-E(k))/\hbar]\\
=\frac{\pi m}{\hbar^2 k}\left[\delta(k^\prime-k)-\delta(k^\prime+k)\right].
\end{eqnarray*}
Making use a symmetrized expression for the real integral $I$, one can show that the
second term to contain $\delta(k^\prime+k)$ leads to zero input into $I$. As a
result, for the Larmor transmission time, we obtain
\begin{eqnarray*}
\tau^L_{tr}=\frac{m}{{\bf T}\hbar}\int_{-\infty}^{\infty}dk |A_{in}(k)|^2
k^{-1}\int_a^b dx |\psi_{tr}(x,k)|^2.
\end{eqnarray*}
Or, taking into account Exp. (\ref{4005}) as well as the relationship
$\psi_{tr}(x,-k)=\psi_{tr}^*(x,k),$ we eventually obtain that
\begin{eqnarray} \label{823}
\tau^L_{tr}=\frac{1}{{\bf T}}\int_{0}^{\infty}\varpi(k) T(k)\tau^{tr}_{dwell}(k) dk,
\end{eqnarray}
where $\varpi(k)=|A_{in}(k)|^2-|A_{in}(-k)|^2.$

A similar expression takes place for $\tau^L_{ref}$ -
\begin{eqnarray} \label{824}
\tau^L_{ref}=\frac{1}{{\bf R}}\int_{0}^{\infty}\varpi(k) R(k)\tau^{ref}_{dwell}(k)
dk.
\end{eqnarray}
The integrands in both these expressions are evident to be non-singular at $k=0$.

Thus, the Larmor times for transmission and reflection are, like the local dwell
time (see \cite{Ha1,Le1,Ha2}), the average values of the corresponding dwell times.

In the end of this section it is useful again to address rectangular barriers. For
the stationary case, in addition to Larmor times (\ref{4007}), (\ref{4009}),
(\ref{40030}) and (\ref{40031})), we present explicit expressions for the initial
angles $\theta_{tr}^{(0)}$ and $\phi_{tr}^{(0)}$. To the first order in $\omega_L$,
we have $\theta_{tr}^{(0)}=\frac{\pi}{2}-\omega_L \tau_z,$ $\phi_{tr}^{(0)}=\omega_L
\tau_0,$ $\theta_{ref}^{(0)}=\frac{\pi}{2}+\omega_L \tau_z$ and
$\phi_{tr}^{(0)}=-\omega_L \tau_0,$ where
\begin{eqnarray*}
\tau_z=\frac{m\kappa_0^2}{\hbar\kappa^2}\cdot\frac{(\kappa^2-k^2)\sinh(\kappa
d)+\kappa^2_0\kappa d\cosh(\kappa d)}{4k^2\kappa^2+\kappa_0^4\sinh^2(\kappa d)}
\sinh(\kappa d)\\
\tau_z=\frac{m\kappa_0^2}{\hbar\kappa^2}\cdot\frac{\kappa^2_0\kappa d\cos(\kappa
d)-\beta(\kappa^2+k^2)\sin(\kappa d)}{4k^2\kappa^2+\kappa_0^4\sin^2(\kappa d)}
\sin(\kappa d),
\end{eqnarray*}
for $E<V_0$ and $E\geq V_0$, respectively;
\begin{eqnarray} \label{90028}
\tau_0=\frac{2mk}{\hbar\kappa}\cdot\frac{(\kappa^2-k^2)\sinh(\kappa
d)+\kappa^2_0\kappa d\cosh(\kappa d)}{4k^2\kappa^2+\kappa_0^4\sinh^2(\kappa d)},\nonumber\\
\tau_0=\frac{2mk}{\hbar\kappa}\cdot\frac{\beta\kappa^2_0\kappa d\cos(\kappa
d)-(\kappa^2+k^2)\sin(\kappa d)}{4k^2\kappa^2+\kappa_0^4\sin^2(\kappa d)},
\end{eqnarray}
for $E<V_0$ and $E\geq V_0$, respectively.

Note that $\tau_z$ is just the characteristic time introduced in \cite{But} (see
Exp. (2.20a)). However, we have to stress once more that this quantity does not
describe the duration of the scattering process (see the end of Section \ref{a332}).
As regards $\tau_0,$ this quantity is directly associated with timing a particle in
the barrier region. It describes the initial position of the "clock-pointers", which
they have before entering this region.

\subsection{Tunneling a particle through an opaque rectangular barrier}\label{a4}

Let us now show that the case of tunneling a particle, with a well defined energy,
through an opaque rectangular potential barrier is the most suitable one to verify
our approach. Let us denote the measured azimuthal angle as $\phi_{tr}^{(\infty)}.$
By our approach $\phi_{tr}^{(\infty)}=\phi_{tr}^{(0)}+\Delta\phi_{tr}$. That is, the
final time to be registered by the particle's "clocks" should be equal to
$\tau_0+\tau^L_{tr}.$

As is seen, in the general case there is a problem to distinguish the inputs
$\tau_0$ and $\tau^L_{tr}.$ However, for a particle tunneling through an opaque
rectangular barrier this problem disappears. The point is that for $\kappa d\gg 1,$
$|\tau_0|\ll\tau^L_{tr}$ (see Exps. (\ref{4007}) and (\ref{90028})).

Note, in the case considered, Smith's dwell time ($\tau^{Smith}_{dwell}$), which
coincides with the "phase" time, and Buttiker's dwell time (see Exps. (3.2) and
(2.20b) in \cite{But}) saturate with increasing the barrier's width. Just this
property of the tunneling times is interpreted as the Hartman effect. At the same
time, our approach denies the existence of the Hartman effect: transmission time
(\ref{4007}) increases exponentially when $d\to\infty.$

Note that the Bohmian approach formally denies this effect, too. It predicts that
the time, $\tau_{dwell}^{Bohm},$ spent by a transmitted particle in the opaque
rectangular barrier is
\begin{eqnarray*}
\tau^{Bohm}_{dwell}\equiv\frac{1}{T}\tau^{Smith}_{dwell}=\frac{m}{2\hbar
k^3\kappa^3}\Big[\left(\kappa^2-k^2\right)k^2\kappa d\\ +\kappa_0^4 \sinh(2\kappa
d)/2\Big].
\end{eqnarray*}
Thus, for $\kappa d\gg 1$ we have $\tau^{Bohm}_{dwell}/\tau^{tr}_{dwell}\sim
\cosh(\kappa d),$ i.e., \[\tau^{Bohm}_{dwell}\gg\tau^{tr}_{dwell}\gg
\tau^{Smith}_{dwell} \sim \tau^{Butt}_{dwell}.\]

As is seen, in comparison with our definition, $\tau^{Bohm}_{dwell}$ overestimates
the duration of dwelling transmitted particles in the barrier region. In the final
analysis, this sharp difference between $\tau^{Bohm}_{dwell}$ and
$\tau^{tr}_{dwell}$ is explained by the fact that $\tau_{dwell}^{Bohm}$ to describe
transmission was obtained in terms of $\psi_{full}.$ One can show that the input of
the to-be-reflected subensemble of particles into $\int_a^b|\psi_{full}(x,k)|^2 dx$
dominates inside the region of an opaque potential barrier. Therefore treating this
time scale as a characteristic time for transmission has no basis.

As was said (see Sections \ref{aI} and \ref{a0}), the trajectories of transmitted
and reflected particles are ill-defined in the Bohmian mechanics. However, we have
to stress that our approach does not at all deny the Bohmian one. It suggests only
that causal trajectories for these particles should be redefined. An incident
particle should have two possibility (to be transmitted or to reflected by the
barrier) irrespective of the location of its starting point. This means that just
two causal trajectories should evolve from each staring point: on the $OX$-axis one
should lead to plus infinity, but another should approach minus infinity. Both sets
of causal trajectories must be defined on the basis of $\psi_{tr}(x,t)$ and
$\psi_{ref}(x,t).$ As to the rest, all mathematical tools developed in the Bohmian
mechanics (see, e.g., \cite{Bo1,Kr1}) remain in force.

Tunneling is useful also to display explicitly the role of the exact and asymptotic
group times. Fig.1 shows the time dependence of the mean value of the particle's
position for transmission, where $a=200nm$, $b=215nm$, $V_0=0.2 eV$. At $t=0$ the
(full) state of the particle is described by the Gaussian wave packet peaked around
$x=0$; its half-width $10nm$; the average energy of the particle $0.05eV$.

As is seen, the exact group time gives the time spent by the CM of the transmitted
wave packet in the barrier region. But the asymptotic time displays its lag, long
after the scattering event, with respect to the CM of a packet, to start from the
point $x_{start}^{tr}$ and move freely with the velocity $\hbar <k>^{tr}_{out}/m$.

In this case the exact group transmission time is equal approximately to $0.155ps$,
the asymptotic one is of $0.01ps$, and $\tau_{free}\approx 0.025ps$. As is seen, the
dwell and exact group times for transmission, both evidence that, though the
asymptotic group time for transmission is small for this case, transmitted particles
spend much time in the barrier region. Note, also that the times spent by
transmitted and reflected particles in the barrier region do not coincide even for
symmetric barriers.

\section{Conclusion}

It is shown that a 1D completed scattering is a combination of two sub-processes,
transmission and reflection, evolved coherently. In the case of symmetric potential
barrier we find explicitly two solutions to the Schr\"odinger equation, which
describe these sub-processes at all stages of scattering. Their sum gives the wave
function to describe the whole combined process.

On the basis of these solutions, for either sub-process, we define the time spent,
on the average, by a particle in the barrier region. For this purpose we reconsider
the well-known group, dwell and Larmor-time concepts. The group time concept is
suitable for timing a particle in a well-localized state, when the width of a wave
packet is smaller than the barrier's width. The dwell time concept is introduced for
timing a particle in the stationary state. The Larmor "clock" is the most universal
instrument for timing the motion of transmitted and reflected particles, without
influence on the scattering event. It is applicable for any wave packets. We found
that the Larmor times for transmission and reflection are the average values of the
corresponding dwell times. The results of our theory can be verified experimentally.

\section*{Figure captions}
%\begin{verbatim}

%\begin{verbatim}
Fig.1 The $t$-dependence of the average position of transmitted particles (solid
line); the initial (full) state vector represents the Gaussian wave packet peaked
around the point $x=0$, its half-width equals to $10nm,$ the average kinetic
particle's energy is $0.05eV;$ $a=200nm$, $b=215nm$.
%\end{verbatim}

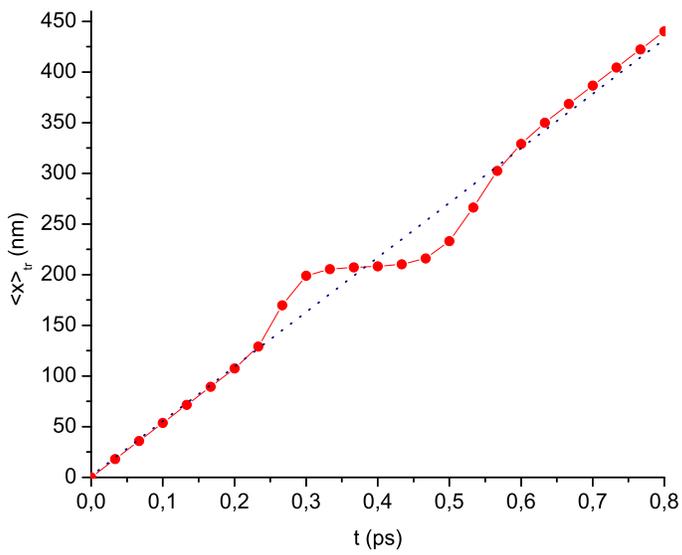
\end{document}